\documentclass[twocolumn,prb,floatfix,nobibnotes]{revtex4-2}
\usepackage[utf8]{inputenc}
\usepackage{graphicx}
\usepackage{amsmath}
\usepackage[colorlinks=true,citecolor=blue]{hyperref}
\usepackage{braket}

\begin{document}
\title{Phonon- and magnon-mediated decoherence of a magnonic qubit}
\author{Vemund Falch} 
\author{Arne Brataas}
\author{Jeroen Danon}
\affiliation{Center for Quantum Spintronics, Department of Physics, Norwegian University of Science and Technology, NO-7491 Trondheim, Norway}
\date{October 16, 2025}

\newcommand{\td}{{\text{d}}}
\newcommand{\Tr}{{\text{Tr}}}
\newcommand{\hH}{{\hat{H}}}
\newcommand{\hV}{{\hat{V}}}
\newcommand{\bp}{{\boldsymbol{p}}}
\newcommand{\bR}{{\boldsymbol{R}}}
\newcommand{\br}{{\boldsymbol{r}}}
\newcommand{\hR}{{\hat{R}}}
\newcommand{\bq}{{\boldsymbol{q}}}
\newcommand{\bk}{{\boldsymbol{k}}}
\newcommand{\ha}{{\hat{a}}}
\newcommand{\hb}{{\hat{b}}}
\newcommand{\hX}{{\hat{\xi}}}
\newcommand{\hv}{{\hat{v}}}
\newcommand{\hrho}{{\hat{\rho}}}
\newcommand{\bPhi}{{\bar{\Phi}}} 
\newcommand{\tq}{{\text{q}}} 
\newcommand{\tc}{{\text{c}}} 
\newcommand{\bS}{{\boldsymbol{S}}} 
\newcommand{\be}{{\boldsymbol{e}}}

\begin{abstract}
We investigate the decoherence of magnonic qubits in small ferromagnetic insulators and compute the relaxation and dephasing rates due to magnon-phonon and magnon-magnon interactions. We combine a Bloch--Redfield description with Keldysh non-equilibrium field theory to find explicit expressions for the rates. For a quadratic dispersion and assuming a uniform mode defines the qubit, we find that decay into two phonons is the only allowed relaxation process at zero temperature. The low resonance frequency and heavy unit cell strongly suppress this process in yttrium-iron-garnet. We also find that the dephasing rate scales with the inverse of size and damping of the magnet, and could become large for small and clean magnets. Our calculation thus provides additional insight into the viability of magnon-based quantum devices.
\end{abstract}

\maketitle

\section{\label{sec:Introduction}Introduction}

The field of quantum computation \cite{Feynman1982, QCReview:Nature2010} has advanced significantly over the past few decades, culminating in recent claims of quantum advantage, where quantum processors outperform the most powerful classical computers for specific tasks \cite{Google2019Supremacy, Kim2023, king2024QunatumSupremacy}. A variety of platforms for realizing qubits, the fundamental units of quantum information, have been proposed, each offering distinct advantages \cite{SpinQubitReview, SemiconductorQubitsReview2, TrappedAtomsReview, NeutralAtomsQuBit, SCRev, TrappedIonReview}. However, achieving an optimal combination of long coherence times, high gate fidelities, fast operation speeds, and scalability sufficient for practical applications remains an outstanding challenge \cite{Kim2023, QCPrespective:NatureCom, Ganjam2024:ErrorRates, TrappedIonReview, QuBitMaterial:Review}. As such, the optimal implementation of qubits remains an open question.

Magnons, quantized spin excitations in ordered magnets, have attracted considerable interest due to their fundamental properties and potential applications \cite{YIGReview,CavityMagnonicsReview,QuantumMagnonics}. Ferromagnets have already been employed in conventional memory technologies \cite{STTMRAM,SOTMRAM}, and ongoing research explores their use in spin-wave-based computing architectures \cite{SpinWaveComputing,NanomagnetReversalMemory,NanomagnetReversalMemory_2}. Over the past decade, significant effort has also been devoted to uncovering the non-classical behavior of magnons, including the realization of single-magnon states \cite{SingleMagnonStateObservation,SingleMagnonDetection}, antibunched magnons \cite{AntiBunchedExtreme,AntiBunching}, magnon bundles \cite{MagnonBundle,MagnonPhotonBundle}, squeezed states \cite{SqueezedMagnons,MagnonSqueezing}, cat states \cite{MagnonCatState,MagnonCatState2,MagnonCatState3}, and entangled magnons \cite{MagnonEntanglementKerr,MagMagEntanglement2,MagPhotPhonEntanglement}. A significant advantage of insulating magnets, such as yttrium iron garnet (YIG), lies in their intrinsically low damping rates \cite{YIGGilbertDampingSphere1,YIGGilbertDampingFilm1,YIGMilliKelvin1,YIGMilliKelvin2}, which have been extensively investigated using various theoretical models \cite{Streib,YIGDispAndInteractions,DampingModelLattice1,FirstPrinciples}. Furthermore, magnons exhibit strong coupling to a variety of quantum systems, including electronic degrees of freedom \cite{MagnonSpintronics,SpinInsulatronics:Brataas2020}, photons \cite{MagnonTravellingWaveInteraction,MagPhotCoupl,MagPhotCoupl2,MagPHotCoupl3}, superconducting qubits \cite{MAGSCQubitCoupl,MagQubit_DoubleCav,MagQubitCouplDirect}, and phonons \cite{MagPhonCoupl,Streib}. Non-Hermitian engineering of magnon interactions has also been proposed as a route to improved control \cite{DissipativeCouplMagPhot,NonHermitianCoupl:Kim2024}. In parallel, recent advances have demonstrated high-sensitivity magnon detection techniques, reaching the level of single-magnon sensitivity \cite{MagnonSensing,MagnonSensing:PRAppl,SingleMagnonDetection}.

Several proposals have explored the use of magnons for quantum information processing \cite{QuantumMagnonics}, either by integrating magnonic systems with existing quantum computing platforms as logic gates \cite{MagnetQubitGate,MagnonGates2,MagQubitCouplDirect} or as quantum memories \cite{MagQubit_DoubleCav}. More recently, magnonic systems were proposed as platforms for hosting qubits. One approach leverages the double-well structure formed by two magnons due to dipolar interactions \cite{MagnonBEC_TwoWellQB}. Another proposal isolates a qubit within a single magnonic mode by exploiting the nonlinear magnon Kerr effect \cite{YuanBlockade}, which induces an anharmonic energy level structure, enabling the magnon mode to operate as an effective two-level system in analogy with superconducting qubits \cite{QCReview:Nature2010}. A complete set of quantum logic gates for such a magnon qubit was outlined in Ref.~\cite{MagnonQubit}. 

In this work, we focus on the coherence properties of a magnon-based two-level system, which couples to both phonons and other magnons. To compute its decoherence rates, we combine the Bloch--Redfield master equation approach with the Keldysh formalism \cite{Keldysh_Kamenev}, which incorporates dissipation effects in the environment \cite{KeldyshwDamp1,KeldyshDamp2}. We calculate analytical expressions for the decoherence rates of the uniform Kittel mode for a simple magnon model with quadratic dispersion, neglecting dipolar interactions. We find that only two-phonon-mediated spontaneous decay survives in the zero-temperature limit, but is strongly suppressed in YIG, due to the low uniform resonance frequency and large, and thus heavy, unit cell. For pure dephasing processes, we find that the dephasing rate is inversely proportional to both the Gilbert damping and the size of the magnet. It can thus be large for a small and clean magnet, although the pure dephasing rate also vanishes in the zero-temperature limit. For large magnets, dephasing due to momentum non-conserving processes \cite{YuanDephasing,MagnonQubit} is instead expected to dominate.

We organize our presentation as follows. Section \ref{sec:MasterEq} introduces the Hamiltonian of the system, coupling the magnons and phonons. In Sec.~\ref{sec:BlochRed} we derive the master equation to find the decoherence rates. We calculate the relaxation rate in Sec.~\ref{sec:rel} and the dephasing rate in Sec.~\ref{sec:deph}, before presenting our conclusions in Sec.~\ref{sec:Concs}.

\section{\label{sec:MasterEq}System Hamiltonian}

\begin{figure}[t]
    \centering 
    \includegraphics[width=\linewidth]{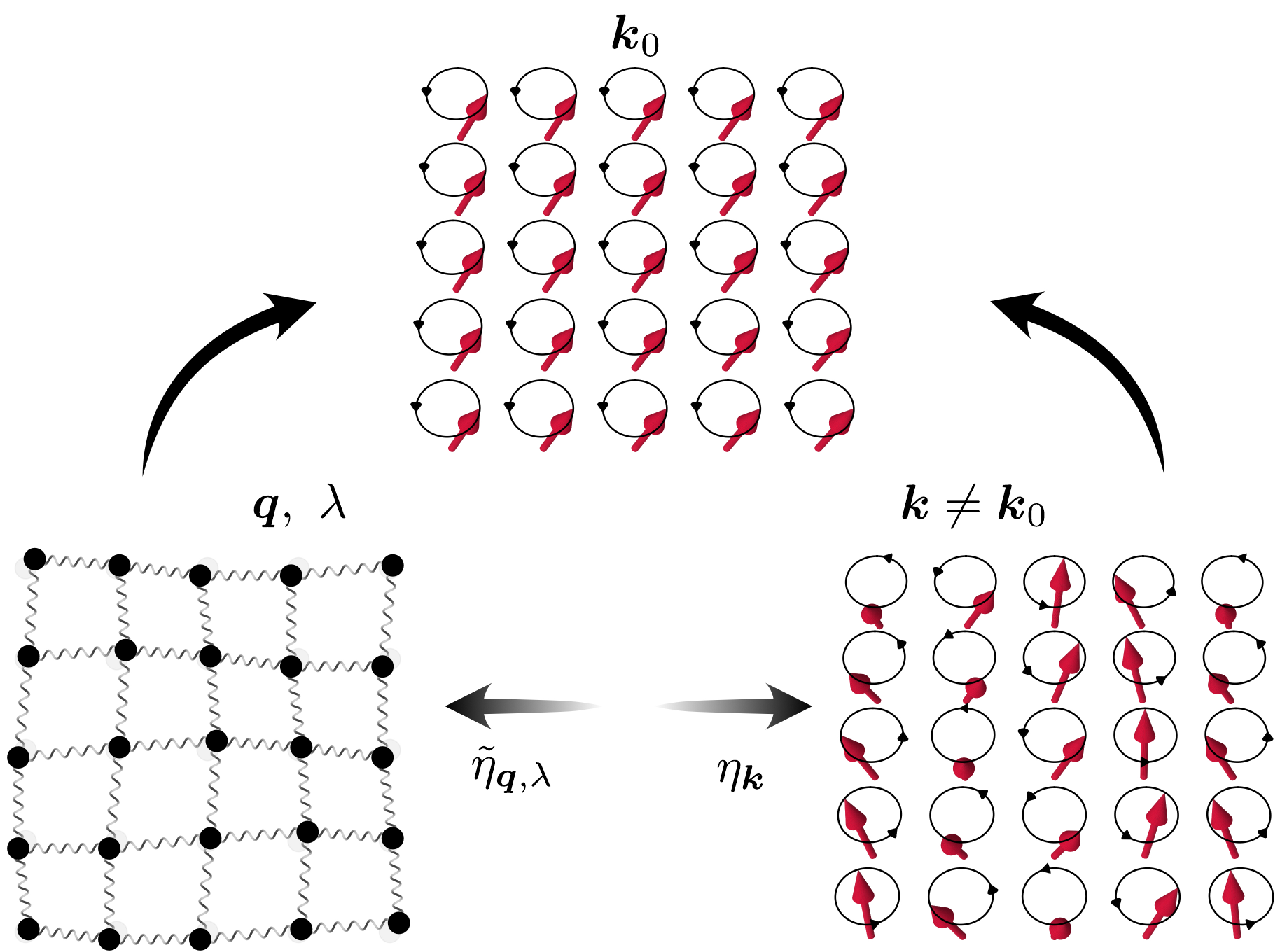}
    \caption{The system under consideration. A specific magnon mode with wave vector $\bk_0$ is coupled to an environment of other magnons with wave vectors $\bk\neq\bk_0$ and phonon modes with wave vectors $\bq$ and polarization $\lambda$. The coupling of the environment to itself is encoded in a phenomenological damping rate $\eta_\bk$ ($\tilde{\eta}_{\bq,\lambda}$) for the environmental magnons (phonons). }
    \label{fig:CouplDiag}
\end{figure}

We consider an insulating nanomagnet, which hosts interacting magnon and phonon modes, as illustrated in Fig.~\ref{fig:CouplDiag}. The nanomagnet is assumed to be sufficiently small so that the energy separation between different magnon modes is significant compared to their linewidths, allowing selective pumping and probing of a single magnon mode of interest, characterized by its wave vector $\bk_0$~\cite{YuanBlockade}. Moreover, due to the magnon Kerr effect, the energy level spacing within this $\bk_0$ mode becomes nonuniform, enabling the isolation of two specific states, $\ket{n_{\bk_0}}$ and $\ket{m_{\bk_0}}$, to form an effective two-level system and operate as a qubit~\cite{QCReview:Nature2010, MagnonQubit}. However, the interactions with other magnon modes and phonons will lead to decoherence and the degradation of the qubit's coherence. This work investigates the relaxation and dephasing rates of the $\bk_0$ magnon mode arising from its interactions with the surrounding magnonic and phononic environments.

To analyze the system, we partition it into two subsystems: a specific magnon mode at wave vector $\bk_0$ and its environment $E$, consisting of all other magnon and phonon modes. The total Hamiltonian is written as $\hH = \hH_0 + \hH_E + \hV$, where $\hH_0$ and $\hH_E$ describe the $\bk_0$ magnon mode and the environmental modes, respectively, each as a collection of harmonic oscillators, 
\begin{subequations}
    \begin{align}
        \hH_0&=\hbar\omega_{\bk_0}\hb^\dagger_{\bk_0}\hb_{\bk_0}^{\phantom{\dagger}},\\
        \hH_E&=\sum_{\bk\neq\bk_0} \hbar \omega_\bk\hb^\dagger_{\bk}\hb_{\bk}^{\phantom{\dagger}} + \sum_{\bq,\lambda}\hbar \tilde{\omega}_{\bq,\lambda}\ha^\dagger_{\bq,\lambda}\ha_{\bq,\lambda}^{\phantom{\dagger}},\label{eq:HE_0}
    \end{align}
\end{subequations}
where $\hb_\bk^\dagger$ and $\ha_{\bq,\lambda}^\dagger$ are magnon and phonon annihilation (creation) operators for modes $\bk$ and $(\bq,\lambda)$, respectively, with corresponding frequencies $\omega_\bk$ and $\tilde{\omega}_{\bq,\lambda}$, where $\lambda$ labels the polarization of the phonon. The interaction term $\hV = \hV^{(p)} + \hV^{(m)}$ contains two contributions: $\hV^{(p)}$, describing the coupling between the $\bk_0$ magnons and the phonon modes, and $\hV^{(m)}$, describing the interaction between the $\bk_0$ magnons and the other magnon modes. Internal interactions among the environmental modes are not treated explicitly; instead, their effects are incorporated phenomenologically through an effective damping rate for the environmental modes.

Magnon--phonon coupling in insulating ferromagnets, particularly in the prototypical material YIG, has been extensively studied~\cite{Streib,YIGDispAndInteractions}.
For our purposes, the dominant contributions to the coupling can be expressed as $\hV^{(p)} = \hV^{(11)} + \hV^{(12)} + \hV^{(21)}$, with
\begin{subequations}
\begin{align}
	\hV^{(11)}= {} & {} \hb_{\bk_0}^\dagger\sum_{\lambda}B^{(11)}_{\lambda}\hX_{\bk_0,\lambda}+\text{H.c.},\\
	\hV^{(12)} = {} & {} \hb_{\bk_0}^\dagger\sum_{\bq,\lambda_1,\lambda_2}B^{(12)}_{\bq,\lambda_1,\lambda_2} \hX_{\bq,\lambda_1}\hX_{\bk_0-\bq,\lambda_2}+\text{H.c.},\\
        \begin{split}\hV^{(21)} = {} & {} \hb_{\bk_0}^\dagger\sum_{\bq,\lambda} B^{(21)}_{\bq,\lambda}\hb_{\bk_0-\bq}\hX_{\bq,\lambda} \\
        {} & {} +\hb_{\bk_0}^\dagger\sum_{\bq,\lambda}\tilde{B}^{(21)}_{\bq,\lambda}\hb_{\bq-\bk_0}^\dagger\hX_{\bq,\lambda} + \text{H.c.},\end{split}
\end{align}\label{eq:MagPhonInteraction}%
\end{subequations}
where we use the notation $\hX_{\bq,\lambda}=\ha_{\bq,\lambda}+\ha^\dagger_{-\bq,\lambda}$, which is proportional to the phononic displacement operator.
The matrix elements $B^{(nm)}$, given explicitly in Appendix~\ref{secA:Hams} for a cubic crystal, characterize the amplitudes of scattering processes involving $n$ magnons and $m$ phonons, with the creation or annihilation of one magnon in the $\bk_0$ mode.

Due to conservation of energy and momentum, the one-magnon--one-phonon scattering process described by $\hV^{(11)}$ is resonant only when a phonon exists with both wave vector $\bk_0$ and energy $\hbar\tilde{\omega}_{\bk_0,\lambda} \approx \hbar\omega_{\bk_0}$. In this work, we focus on the regime where $\bk_0$ is sufficiently detuned from the magnon--phonon crossing points, allowing us to neglect $\hV^{(11)}$ and consider only the terms $\hV^{(p)} = \hV^{(12)} + \hV^{(21)}$ in the interaction.

To describe the magnon-magnon interaction, we consider a simple ferromagnet model with only exchange coupling and easy-axis anisotropy, such that we obtain
\begin{align}
        \hV^{(m)} &=\hb_{\bk_0}^\dagger\sum_{\bk,\bk'\neq \bk_0} B^{(m)}_{\bk,\bk'}\hb_{\bk}^\dagger\hb_{\bk'}\hb_{\bk_0+\bk-\bk'}+\text{H.c}\nonumber\\
        &\phantom{=} +\hb_{\bk_0}^\dagger\hb_{\bk_0}\sum_{\bk \neq \bk_0}\tilde{B}^{(m)}_{\bk}\hb_{\bk}^\dagger\hb_{\bk},\label{eq:Mag4Int}
\end{align}
where we again singled out the terms involving the mode $\bk_0$.
We give the resulting explicit expressions for the energies $B^{(m)}$ and $\tilde{B}^{(m)}$ in Appendix \ref{secA:Hams}. Including dipolar interactions would also give cubic interactions \cite{YIG_FilmDisp,YIGDispAndInteractions}.

\section{\label{sec:BlochRed}Bloch--Redfield equations}
We aim to describe the dissipative dynamics of the magnon qubit arising from its interactions with the environment. We begin with the Nakajima--Zwanzig master equation, which governs the time evolution of the reduced density matrix $\hrho_{0}(t) = \Tr_E{\hrho(t)}$ for the $\bk_0$ magnon mode. Assuming that the environment is initially in thermal equilibrium and uncorrelated with the $\bk_0$ mode, that is, $\hrho(0) = \hrho_{0}(0) \otimes \hrho_E^{\rm th}$ with $\hrho_E^{\rm th} \propto e^{-\hH_E/T}$ setting $k_{\rm B}=1$, the Redfield equation for $\hrho_0(t)$ can be derived within the Born--Markov approximation~\cite{Bloch_Nakajima_Zwansig}
\begin{equation}
    \frac{\td}{\td t}\rho_{0}^{nm}=-i\omega_{nm}\rho_{0}^{nm}-\sum_{kl}R_{nmkl}\rho_{0}^{kl},\label{eq:RedfieldEQ}
\end{equation}
for the matrix elements $\rho_{0}^{nm}=\braket{n\vert\hrho_{0}\vert m}$ where $\ket{n}$ and $\ket{m}$ are $\bk_0$-magnon number eigenstates, i.e., $\hb_{\bk_0}^\dagger\hb_{\bk_0}\ket{n} = n\ket{n}$, and we use the notation $\omega_{nm}=(n-m)\omega_{\bk_0}$.
The tensor $R$ is defined by~\cite{RedfieldSecular,RedfieldSecular2}
\begin{align}
    R_{nmkl}= {} & {} \sum_\alpha \big(\delta_{l,m}\Gamma_{n\alpha\alpha k}+\delta_{k,n}\Gamma^{*}_{m\alpha \alpha l}\big) \nonumber\\
    {} & {} -\Gamma_{lmnk}-\Gamma^*_{knml},\label{eq:RedfieldtensorComplex}
\end{align}
in terms of the correlation functions
\begin{align}
    \Gamma_{nmkl}=  {} & {} \int_0^\infty \frac{\td t}{\hbar^2}  \,e^{-i\omega_{kl}t} 
    \big\langle \hv_{nm}(t)\hv_{kl}(0) \big\rangle.
\label{eq:RedfieldWeight}
\end{align}
where $\hv_{nm}(t) = \braket{n|e^{\frac{i}{\hbar} \hH_E t} \hV e^{-\frac{i}{\hbar} \hH_E t}|m}$ and taking the average amounts to tracing over the environmental degrees of freedom, $\langle \dots \rangle = \Tr_E \{\dots \hat{\rho}_E^{\rm th}\}$.

We now focus on the two qubit-states $\ket{0}$ and $\ket{1}$, the two lowest-occupation states of the $\bk_0$ mode, for simplicity.
Writing the $2\times 2$ qubit density matrix $\hat\rho_0$ in terms of a Bloch vector, $\bp = {\rm Tr}\{ \boldsymbol\sigma \hat\rho_0 \}$, with $\boldsymbol\sigma$ being the vector of the three Pauli matrices, one can recast Eq.~(\ref{eq:RedfieldEQ}) in the form
\begin{equation}
\frac{\td \bp}{\td t}= -\boldsymbol{\omega}\times\bp-\hR(\bp-\bp_0)\label{eq:RedfieldBloch},
\end{equation}
describing the precession and relaxation of the state vector $\bp$.
Within a rotating-wave approximation, i.e., neglecting all rapidly oscillating contributions to Eq.~(\ref{eq:RedfieldEQ}) by keeping only the $R_{nmkl}$ with $n-m = k-l$, one finds for the qubit precession frequency $\boldsymbol{\omega} = (\omega_{\bk_0}+\text{Im}\{R_{1010}\})\hat z$ and that relaxation of $\bp$ to the equilibrium state $\bp_0=(R_{1111}-R_{0000})/(R_{1111}+R_{0000})\hat{{z}}$ is described by the matrix $\hat R = {\rm diag}\{ \frac{1}{2}\Gamma_1 + \Gamma_\phi,  \frac{1}{2}\Gamma_1 + \Gamma_\phi, \Gamma_1 \}$, where
\begin{subequations}\label{eq:TDecor}
    \begin{align}
    \Gamma_1 &=\int \! \frac{\td t}{\hbar^2} \, e^{i\omega_{\bk_0}t} \big\langle \{ \hv_{10}(t) , \hv_{01}(0)\} \big \rangle, \label{eq:T1ExpressionCorr}\\
    \Gamma_\phi &=\int \!\! \frac{\td t}{4\hbar^2} \,  \big\langle \{ [ \hv_{11}(t)- \hv_{00}(t) ],[ \hv_{11}(0)- \hv_{00}(0) ] \}\big\rangle,\label{eq:T2ExpressionCorr}
\end{align}
\end{subequations}
with $\{A,B\} = AB + BA$ denoting the anticommutator.
The relaxation rate $\Gamma_1$ accounts for the effect of inelastic transitions between the qubit states~\footnote{We have neglected scattering into higher levels $\ket{n}$ with $n>1$, which would also contribute to relaxation. A straightforward generalization of the results to also include the state $\ket{2}$ yields the correction $\text{Re}\{\Gamma_{0220}+\Gamma_{1221}\}$ to $\Gamma_1$, where $\Gamma_{0220}$ and $\Gamma_{1221}$ describe exciting the qubit from $\ket{0}$ to $\ket{1}$ and $\ket{1}$ to $\ket{2}$, respectively. For the interactions under consideration [see Eqs.~\eqref{eq:MagPhonInteraction} and \eqref{eq:Mag4Int}], it follows that $\Gamma_{0220}$ is vanishingly small and $\Gamma_{1221} \approx 2\Gamma_{0110}$. At the low temperatures required to have a well-defined qubit, $k_\text{B}T \ll \hbar\omega_{\bk_0}$, qubit decay processes dominate, and $\Gamma_{1221}$ presents only a minor quantitative correction to the rate.} and the dephasing rate $\Gamma_\phi$ for the cumulative effect of fluctuations of the qubit splitting.
In writing (\ref{eq:TDecor}) we also used the fact that all correlations depend only on time differences, i.e., $\langle \hv_{nm}(t+\tau)\hv_{kl}(\tau)\rangle=\langle \hv_{nm}(t)\hv_{kl}(0)\rangle$, as the environment is in equilibrium.
We further note that in the derivation of the equation of motion above, it is implicitly assumed that the initial density-matrix $\hat{\rho}_E(0)$ of the environment commutes with $\hH_E$, i.e., $[\hat{H}_E,\hat{\rho}_E(0)]=0$, and that the interaction Hamiltonian is traceless, $\Tr_E\{\hV\hrho_E(0)\}=0$. The former assumption holds as we assumed $\hrho_E(0)=\hrho_E^\text{th}$, while for many terms with a finite trace, such as those in Eq.~\eqref{eq:T2ExpressionCorr}, the latter assumption can then be satisfied by instead applying the above formalism to the modified partition $\hV\to\hV-\langle \hV\rangle$ and $\hH_0\to\hH_0+\langle \hV\rangle$, which we implicitly assume in the following. This partition incorporates the first-order corrections due to the coupling to the environment as a renormalization of the parameters in $\hat H_0$, which amounts to calculating only the connected part of the two-point correlation functions in Eq.~\eqref{eq:RedfieldWeight}. One should however be aware that this redefinition breaks down if the environmental Hamiltonian and $\hat \rho_E(0)$ do not commute, which can happen when the environment is explicitly time-dependent or out of thermal equilibrium.

\begin{figure}
    \centering
    \includegraphics[width=\linewidth]{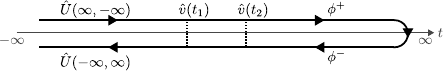}
    \caption{The Keldysh contour. Two operators, $\hat{v}_1$ and $\hat{v}_2$, will be ordered differently on the forward and backward branches.}
    \label{fig:KeldyshContour}
\end{figure}

Assuming that the environment is initially in thermal equilibrium, the correlation functions appearing in Eq.~\eqref{eq:TDecor} can be factorized into products of equilibrium bosonic Green's functions. To compute these expressions explicitly, we employ the Keldysh path integral formalism for nonequilibrium systems \cite{TakeiKeldysh, Keldysh_Kamenev}. The partition function of the environment can be written as 
\begin{equation} 
Z = \frac{\Tr {\hat{U}(-\infty,\infty)\hat{U}(\infty,-\infty)\hat{\rho}_E(-\infty)}}{\Tr{\hat{\rho}_E(-\infty)}}, 
\end{equation} where $\hat{U}(t,t')$ is the time-evolution operator from $t'$ to $t$ and $\hrho_E(t)$ denotes the environmental density matrix, whose time evolution is traced along the Keldysh contour shown in Fig.~\ref{fig:KeldyshContour}. The partition function can be expressed as a functional integral, $Z = \int D[\bar{\Phi},\Phi]\exp(iS)$, over the bosonic fields $\Phi = \begin{bmatrix} \phi & \varphi \end{bmatrix}$ and their complex conjugates $\bPhi$. Here, $\phi = \begin{bmatrix} \phi_\bk \end{bmatrix}$ and $\varphi = \begin{bmatrix} \varphi_{\bq\lambda} \end{bmatrix}$ describe the magnon and phonon fields, respectively. The action $S = S[\bPhi,\Phi]$ fully captures the environment dynamics.

Separating the fields on the forward and backward branches of the contour, $\Phi^+(t)$ and $\Phi^-(t)$, and using the Keldysh rotation $\Phi^{\tc(\tq)}= \frac{1}{\sqrt 2}[ \Phi^+(t)\pm\Phi^-(t)]$, one can write the action of the noninteracting bosonic environmental modes, described by $\hat H_E$, as
\begin{align}
    S_0=\iint_{-\infty}^\infty \!\!\! \td t \, \td t' \begin{bmatrix} \bPhi^\tc(t)&\bPhi^\tq(t)\end{bmatrix} \hat{G}^{-1}(t,t')\begin{bmatrix}\Phi^\tc(t')\\\Phi^\tq(t')\end{bmatrix},
\end{align}
with the Green's function matrix
\begin{equation}
    \hat{G}(t,t')=\begin{bmatrix}
        \hat{G}^K(t,t')&\hat{G}^R(t,t')\\\hat{G}^A(t,t')&0
    \end{bmatrix},
\end{equation}
expressed in terms of the advanced, retarded, and Keldysh Green's functions describing the bosonic modes in the environment. In thermal equilibrium, the Green's functions $\hat{G}^{K/R/A}(t,t')$ are diagonal in the quasiparticle basis. The magnonic Green's functions then take the simple form
\begin{subequations}
    \begin{align}
    G^R_{\bk}(\omega)&=\frac{1}{\omega-\omega_\bk+i\eta_\bk}=\big[G^A_\bk(\omega)]^*,\\
    G^K_\bk(\omega)&=\big[2n(\omega)+1\big]\big[G^R_\bk(\omega)-G^A_\bk(\omega)\big],
    \end{align}\label{eq:KeldyshGreenMag}%
\end{subequations}
where $\eta_\bk$ is a positive parameter describing damping due to the coupling of the magnonic modes to their environment, and we used the Bose distribution function $n(\omega)=[\exp(\hbar\omega/T)-1]^{-1}$.
The phononic Green's functions read similarly as
\begin{subequations}\label{eq:KeldyshGreenPhon}
    \begin{align}
    \tilde G^R_{\bq,\lambda}(\omega)&=\frac{1}{\omega-\tilde \omega_{\bq,\lambda}+i\tilde\eta_{\bq,\lambda} }=\big[\tilde G^A_{\bq,\lambda}(\omega)]^*,\\
    \tilde G^K_{\bq,\lambda}(\omega)&=\big[2n(\omega)+1\big]\big[\tilde G^R_{\bq,\lambda}(\omega)-\tilde G^A_{\bq,\lambda}(\omega)\big],\end{align}
\end{subequations}
where we included a similar damping parameter $\tilde\eta_{\bq,\lambda}$.

To compute correlation functions of operators $\hv_i(t)$ (where $i$ will be $00$, $11$, or $10$ below) within the Keldysh formalism, we introduce source terms by modifying the Hamiltonian as $\hH_E \to \hH_E \pm \sum_i V_i(t) \hv_i$, where the plus and minus signs correspond to the forward $(+)$ and backward $(-)$ branches of the Keldysh contour, respectively. In the path integral formalism, this modification leads to an action of the form 
\begin{equation} S = S_0 - \sum_i \int_{-\infty}^{\infty} \td t\, V_i(t) \tilde{v}_i(\bar{\Phi}^\tc, \bar{\Phi}^\tq, \Phi^\tc, \Phi^\tq), 
\end{equation} 
where $\tilde{v}_i$ denotes the normal-ordered operator $\hv_i$ expressed in terms of the environmental bosonic fields. With the source fields introduced, the correlation functions appearing in Eq.~\eqref{eq:TDecor} can be obtained as functional derivatives of the partition function: \begin{equation} \langle \{ \hv_1(t_1), \hv_2(t_2) \} \rangle = \frac{i^2}{2} \frac{\delta^2 Z}{\delta V_1(t_1) \delta V_2(t_2)}\bigg|_{V_{1,2}=0}.\label{eq:KeldyshExp} 
\end{equation} Applying standard rules for Gaussian integrals over bosonic fields, the magnonic contractions follow 
\begin{equation} 
\langle \phi^\alpha_\bk(t) \bar{\phi}^\beta_{\bk'}(t') \rangle = i \delta_{\bk,\bk'} G_\bk^{\alpha\beta}(t,t'), \end{equation} where $\alpha, \beta = (\tc, \tq)$ 
label the classical and quantum Keldysh components, and with the phononic contractions similarly defined. Using these contractions, the correlation functions in Eq.~\eqref{eq:TDecor} can be expressed directly in terms of the equilibrium Green's functions of the environment.

\section{Results} 

\subsection{Relaxation}\label{sec:rel}

\begin{figure}[b!]
    \centering
    \includegraphics[width=\linewidth]{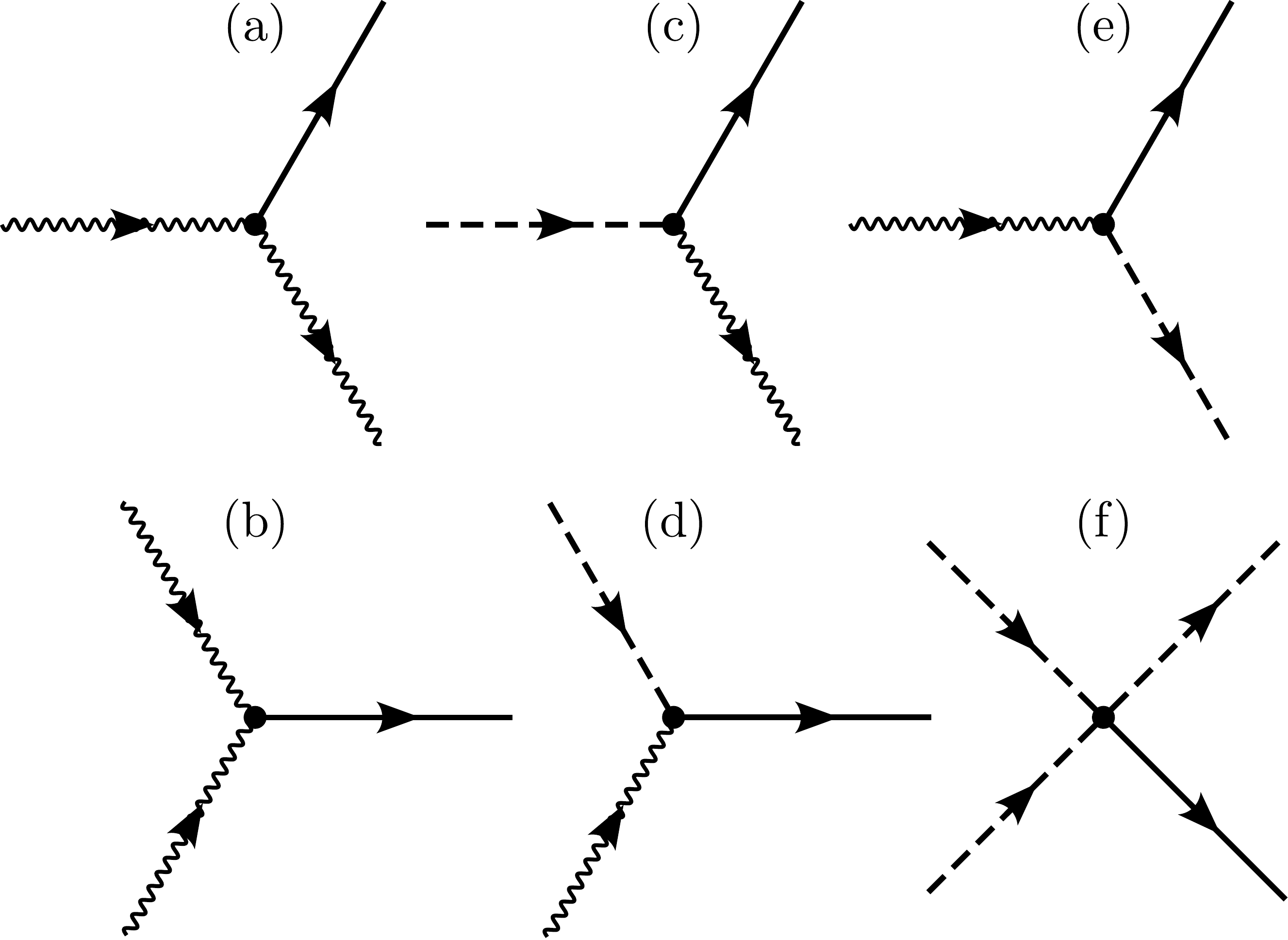}
    \caption{The vertices for the processes contributing to the relaxation rate, where a magnon with specific momentum $\bk_0$ (solid lines) is created due to interactions with phonons (wavy lines) and magnons with wave vectors $\bk\neq\bk_0$ (dashed lines). Only the out-scattering vertices are shown; the full interaction also contains the in-scattering vertices, which can be found by time inversion.}
    \label{fig:ScatteringVertices}
\end{figure}

We first focus on the relaxation rate $\Gamma_1$, which involves the operator $\hat v_{10} = \hat v_{10}^{(12)} + \hat v_{10}^{(21)} +\hat v_{10}^{(m)}$ with individual contributions given by 
\begin{subequations}\label{eq:vsGamma1}
    \begin{align}
    \hat v_{10}^{(12)} &= \sum_{\bq,\lambda_1,\lambda_2}B^{(12)}_{\bq,\lambda_1,\lambda_2} \hX_{\bq,\lambda_1}\hX_{\bk_0-\bq,\lambda_2},\\
    \hat v_{10}^{(21)}  &= \sum_{\bq,\lambda} \left( B^{(21)}_{\bq,\lambda}\hb_{\bk_0-\bq} + \tilde{B}^{(21)}_{\bq,\lambda}\hb_{\bq-\bk_0}^\dagger \right)\hX_{\bq,\lambda},\\
    \hat v_{10}^{(m)}  &= \sum_{\bk,\bk'\neq \bk_0} B^{(m)}_{\bk,\bk'}\hb_{\bk}^\dagger\hb_{\bk'}\hb_{\bk_0+\bk-\bk'},
    \end{align}
\end{subequations}
and $\hat v_{01} = \hat v_{10}^\dagger$, where the out-scattering vertices are shown in Fig.~\ref{fig:ScatteringVertices}.

Due to the structure of the interaction terms, interference cross terms vanish upon thermal averaging, allowing $\Gamma_1$ to be decomposed into four independent contributions. Explicit expressions for these contributions, including the effects of finite magnonic and phononic damping rates, are provided in Eq.~(\ref{eq:relaxationfiniteeps}) in Appendix~\ref{app:explicit}.

In the limit of weak damping, i.e., small $\eta_\bk$ and $\tilde\eta_{\bq,\lambda}$, the relaxation rates simplify considerably, yielding the separate contributions
\begin{widetext}
\begin{subequations}
\begin{align}
\Gamma_{1}^{(12)} = {} & {} \frac{2\pi}{\hbar^2} \!\! \sum_{\bq,\lambda_1,\lambda_2} \!\! \big|B^{(12)}_{\bq,\lambda_1,\lambda_2}\big|^2
\bigg\{ N^+(\tilde{\omega}_{\bq,\lambda_1},\tilde{\omega}_{\bk_0-\bq,\lambda_2}) \Big[ L(z^*_{\bk_0}-\tilde{z}_{\bq,\lambda_1}-\tilde{z}_{\bk_0-\bq,\lambda_2})+L(z^*_{\bk_0}+\tilde{z}^*_{\bq,\lambda_1}+\tilde{z}^*_{\bk_0-\bq,\lambda_2})\Big]\nonumber\\
 {} & {} \hspace{8.5em}+2N^-(\tilde{\omega}_{\bq,\lambda_1},\tilde{\omega}_{\bk_0-\bq,\lambda_2}) L(z^*_{\bk_0}+\tilde{z}^*_{\bq,\lambda_1}-\tilde{z}_{\bk_0-\bq,\lambda_2}) \bigg\},\\
\Gamma_{1}^{(21)} =  {} & {} \frac{\pi}{\hbar^2} \sum_{\bq,\lambda} \big|B^{(21)}_{\bq,\lambda}\big|^2
\bigg\{ N^+(\omega_{\bk_0-\bq},\tilde{\omega}_{\bq,\lambda})L(z^*_{\bk_0}-z_{\bk_0-\bq}-\tilde{z}_{\bq,\lambda}) + N^-(\omega_{\bk_0-\bq},\tilde{\omega}_{\bq,\lambda}) L(z^*_{\bk_0}-z_{\bk_0-\bq}+\tilde z^*_{\bq,\lambda})\bigg\} \nonumber\\
{} & {} + \frac{\pi}{\hbar^2} \sum_{\bq,\lambda} \big| \tilde B^{(21)}_{\bq,\lambda}\big|^2
\bigg\{  N^+(\omega_{\bk_0-\bq},\tilde{\omega}_{\bq,\lambda})L(z^*_{\bk_0}+z^*_{\bk_0-\bq}+\tilde z^*_{\bq,\lambda}) + N^-(\omega_{\bk_0-\bq},\tilde{\omega}_{\bq,\lambda}) L(z^*_{\bk_0}+z^*_{\bk_0-\bq}-\tilde z_{\bq,\lambda})\bigg\},\\
\Gamma_{1}^{(m)} =  {} & {} \frac{4\pi}{\hbar^2} \sum_{\bk,\bk'} \big|B^{(m)}_{\bk,\bk'}\big|^2
    \Big\{ [n(\omega_{\bk}) + 1] n(\omega_{\bk'}) n(\omega_{\bk_0+\bk-\bk'}) \nonumber\\
    {} & {} \hspace{7.5em} + n(\omega_\bk)[n(\omega_{\bk'})+1][n(\omega_{\bk_0+\bk-\bk'})+1] \Big\} L(z^*_{\bk_0}+z^*_{\bk}-z_{\bk'}-z_{\bk_0+\bk-\bk'}),
\end{align}\label{eq:FullGammas}
\end{subequations}
\end{widetext}
where we introduced the notations $z_\bk = \omega_\bk - i \eta_\bk$ and $\tilde z_{\bq,\lambda} = \tilde \omega_{\bq,\lambda} - i\tilde \eta_{\bq,\lambda}$, as well as the functions
\begin{equation}
    N^\pm(\omega_1,\omega_2)= [2n(\omega_1)+1][2n(\omega_2)+1]\pm 1,
\end{equation}
and 
\begin{equation}
    L(z)=\frac{1}{\pi}\frac{{\rm Im}\{z\}}{|z|^2},\label{eq:Lorentzian}
\end{equation}
the latter yielding the Lorentzian functions that describe energy conservation in the presence of finite lifetime broadening.
Note that in the limit of vanishing broadening, $\eta_\bk,\tilde\eta_{\bq,\lambda} \to 0$, one has $L(z) \to \delta({\rm Re}\{z\})$.

By rewriting $N^+(\omega_1,\omega_2) = 2[n(\omega_1)+1][n(\omega_2)+1] + 2n(\omega_1)n(\omega_2)$ and $N^-(\omega_1,\omega_2) = 2[n(\omega_1)+1]n(\omega_2) + 2n(\omega_1)[n(\omega_2)+1]$, it becomes clear that the terms involving $N^\pm$ in $\Gamma_1^{(12)}$ and $\Gamma_1^{(21)}$ represent the sum of the qubit transition rates $\ket{1} \to \ket{0}$ and $\ket{0} \to \ket{1}$, arising from processes that either conserve $(-)$ or do not conserve $(+)$ the total number of bosonic excitations in the environment, with corresponding vertices shown in Figs.~\ref{fig:ScatteringVertices}(a), \ref{fig:ScatteringVertices}(c) and \ref{fig:ScatteringVertices}(e) and Figs.~\ref{fig:ScatteringVertices}(b) and \ref{fig:ScatteringVertices}(d), respectively. The magnon--magnon contribution $\Gamma_1^{(m)}$ similarly contains only non-conserving processes, corresponding to the vertex shown in Fig.~\ref{fig:ScatteringVertices}(f).

We emphasize that the resulting qubit relaxation rate differs from the inverse magnon lifetime as typically calculated, for example, in Ref.~\cite{Streib}, where the focus is on the change in occupation of a particular magnon mode $\bk$. In such treatments, the inverse lifetime depends on the difference between in- and out-scattering rates. Consequently, while in the limit $T \to 0$, where $n(\omega) \to 0$, both approaches yield the same behavior, dominated by quasiparticle-conserving out-scattering processes, the qubit relaxation rate at finite temperatures is necessarily larger than the inverse magnon lifetime.

Finally, we note that the two-magnon--one-phonon contribution $\Gamma_{1}^{(21)}$, which is expected to dominate in room-temperature conditions~\cite{Streib}, was previously considered in Refs.~\cite{Yuan:MasterEquation,MagnonQubit}.

We now calculate explicit relaxation rates, based on the simplified ferromagnetic model for the nanomagnet introduced in Appendix~\ref{secA:Hams}. Specifically, we consider only an external magnetic field, easy-axis anisotropy \footnote{While an easy axis would be barred by symmetry in a cubic crystal, we use it here for simplicity. Including the first symmetry-allowed anisotropy $\propto \sum_\alpha (S_i^\alpha)^4$ instead would give the simple numerical factor $K_z\to9K_z/2$ in the interaction term in Eq.~\eqref{eq:bkex}.}, and exchange interactions, neglecting dipolar interactions. While dipolar effects modify the long-wavelength behavior of the magnon spectrum, they strongly depend on the sample’s size and geometry \cite{YIG_FilmDisp}, and are thus omitted here for simplicity. Under these assumptions and in the long-wavelength limit, the magnon dispersion relation takes a quadratic form,
\begin{equation}
\hbar\omega_\bk=\hbar\omega_0+E_\text{ex}k^2a^2,\label{eq:disp}
\end{equation}
where $\hbar\omega_0=g\mu_{\rm B}B+K_z S$ is the qubit ground-state splitting with $g\mu_{\rm B}B$ being the Zeeman splitting associated with an externally applied magnetic field $B$, $K_z$ being the easy-axis energy, and $S$ being the magnitude of the localized spins; $E_\text{ex}$ is the exchange energy, and $a$ is the lattice constant for a unit cell with mass $m$. Similarly, we assume a linear polarization-dependent phonon dispersion $\hbar\tilde{\omega}_{q\lambda}=\hbar v_\lambda q$, where $v_t$ and $v_l$ are the speed of sound for transverse and longitudinal phonons, respectively.

We choose the qubit mode to be the uniform mode $\bk_0=0$, which can be efficiently coupled to external elements such as microwave cavities~\cite{MagPhotCoupl2} or superconducting qubits~\cite{MAGSCQubitCoupl}, enabling coherent manipulation and readout. In this case, the magnon frequency simplifies to $\hbar\omega_{\bk_0} = \hbar\omega_0$.
We further assume the weak-damping limit, taking $\eta_\bk, \eta_{\bq,\lambda} \to 0$, such that the Lorentzian line shapes reduce to delta functions, as discussed above.
Finally, we assume that the level spacing of the environmental modes is much smaller than the qubit splitting $\hbar\omega_{0}$, which allows us to approximate the momentum sums in Eq.~(\ref{eq:FullGammas}) by integrals, e.g., $\sum_{\bk} \to Na^3(2\pi)^{-3} \int \td^3\bk$, where $N$ is the total number of unit cells~\footnote{Formally, this means that all dimensions must be $\gg 2\pi v_{l,t}/\omega_{0}\sim 200\,\text{nm}$. However, working with discrete levels within the limits of a small magnet would give nongeneralizable results that are heavily size and shape dependent. It should also be noted that several proposals for realizing an effective two-level magnon system that does not require very small magnets exist~\cite{AntiBunchedExtreme,MagBlock2}.}.

We can find analytical expressions for some contributions to the relaxation rate within these assumptions.
For the one-magnon--two-phonon processes, we find
\begin{equation}
    \Gamma_1^{(12)} = 
    \gamma^+_{l,t}+\gamma^-_{l,t} + \gamma^+_{l,l} + \gamma^+_{t,t},
\end{equation}
where 
\begin{align}
    \begin{split}
    \gamma^\pm_{\alpha,\beta}=&\frac{1}{7\pi}\frac{\omega_{0}^4 B_{\alpha,\beta}^2 a^3}{m^2 Sv_{\alpha}v_{\beta}|v_{\alpha}\pm v_{\beta}|^5}N^\pm(\omega_{0}\kappa^\pm_\alpha,\omega_{0}\kappa^\pm_\beta),
    \end{split}\label{eq:2PhonRates}
\end{align}
using the notation $\kappa^\pm_{\alpha,\beta}=v_{\alpha,\beta}/\lvert v_{\alpha}\pm v_{\beta}\rvert$.
The energies $B_{\alpha,\beta}$ are material-dependent magnon--phonon coupling strengths that are given in Eq.~(\ref{eqA:BeffTwoPhonon}) in Appendix~\ref{secA:Hams}.
At low temperatures $T/\hbar \omega_{0} \ll 1$ we find that $N^+ \approx 2$ and $N^- \sim e^{-\hbar\omega_{0}/T} \ll N^+$.
In this limit, we can thus approximate
\begin{align}
    \Gamma_1^{(12)} \approx \frac{1}{112\pi}
    \frac{\omega_{0}^4  a^3}{m^2 S}
    \left[ \frac{B_{l,l}^2}{v_l^7}
    + \frac{B_{t,t}^2}{v_t^7} 
    + \frac{32 B_{l,t}^2}{v_lv_t(v_l+v_t)^5} \right] \!.
\end{align}

For the two-magnon--one-phonon contributions, we find
\begin{equation}
    \Gamma_1^{(21)} = \gamma^{(\text{c})} + \gamma^{(\text{nc})},
\end{equation}
where we distinguished a contribution $(\text{c})$ from processes that conserve the total number of magnons and one $(\text{nc})$ from processes that do not conserve the total number of magnons, given by the vertices in Figs.~\ref{fig:ScatteringVertices}(c) and \ref{fig:ScatteringVertices}(e) respectively.
Explicitly, these two contributions read
\begin{widetext}
\begin{subequations}
\begin{align}
    \gamma^{(\text{c})} = {} & {}
    \frac{1}{10\pi} \frac{\hbar^2 B_\parallel^2 }{ma^3E^3_\text{ex}S^2}
    \big[ 2 v_l N^-(\omega_{0}\!+\Omega_l,\Omega_l)+3 v_t N^-(\omega_{0}\!+\Omega_t,\Omega_t) \big], \\
    \gamma^{(\text{nc})} = {} & {} \frac{1}{10\pi} \frac{\hbar^2 (B_\parallel^2 +B_\perp^2)}{ma^3E^3_\text{ex}S^2}\sum_{s=\pm}
    \bigg[
    \frac{2}{3}\frac{v_l\zeta^3_{l,s}}{\zeta_{l,+}\!-\zeta_{l,-}}
    N^-(\Omega_l \zeta_{l,s}\! - \omega_{0}, \Omega_l \zeta_{l,s})
    + \frac{v_t\zeta^3_{t,s}}{\zeta_{t,+}\!-\zeta_{t,-}}
    N^-(\Omega_t \zeta_{t,s}\!-\omega_{0}, \Omega_t \zeta_{t,s})
    \bigg],
\end{align}
\end{subequations}
\end{widetext}
where we introduced the frequency $\Omega_\lambda = \hbar v_\lambda^2/a^2 E_\text{ex}$, corresponding to the frequency of the scattered phonon in the soft-magnon limit $\omega_0\to 0$, and the dimensionless renormalization factor $\zeta_{\lambda,\pm} = \frac{1}{2}(1 \pm \sqrt{1 - {8}\omega_{0}/\Omega_\lambda})$.
When ${8}\omega_{0}/\Omega_\lambda > 1$ the contribution from the polarization $\lambda$ to $\gamma^{(\text{nc})}$ vanishes.
Note that the exchange contribution introduced in Eq.~(\ref{eq:exchange}) does not contribute in this case, since it is proportional to $\bk_0=0$. At low temperature $N^-(\omega_1,\omega_2) \approx 2[e^{-\hbar\omega_1/T}+e^{-\hbar\omega_2/T}]$ and using the fact that normally $\omega_{0} \ll \Omega_{l,t}$, the dominant contribution comes from the low-frequency $s=-$ branch of $\gamma^{(\text{nc})}$ which gives the approximate expression
\begin{equation}
    \Gamma_1^{(21)} \approx \frac{4}{5\pi} \frac{\hbar^2v_t (B_\parallel^2 +B_\perp^2)}{ma^3E^3_\text{ex}S^2}
    \bigg[
    \frac{2v_l}{3v_t}\frac{\omega_0^3}{\Omega_l^3}+\frac{\omega_0^3}{\Omega_t^3}
    \bigg]e^{-\hbar\omega_0/T}.
\end{equation}

Finally, for the contribution from magnon--magnon interaction, we cannot derive a closed-form expression due to the larger phase space available for scattering;  see Appendix~\ref{app:explicit} for an explicit expression in integral form.
In the low-temperature limit, we can again expand all Bose distributions, allowing us to derive the approximate expression
\begin{equation}
    \Gamma_1^{(m)} \approx \frac{T^2K_z^2}{16 \pi^3 \hbar E_\text{ex}^3}e^{-\hbar\omega_{0}/T},\label{eq:MagnonDecay}
\end{equation}
where the dominant contributions come from modes with momenta up to $ak \sim \sqrt{T/E_\text{ex}}$, implying that this result is valid only for $T \ll E_{\rm ex}$.

\begin{figure}
    \centering
    \includegraphics[width=\linewidth]{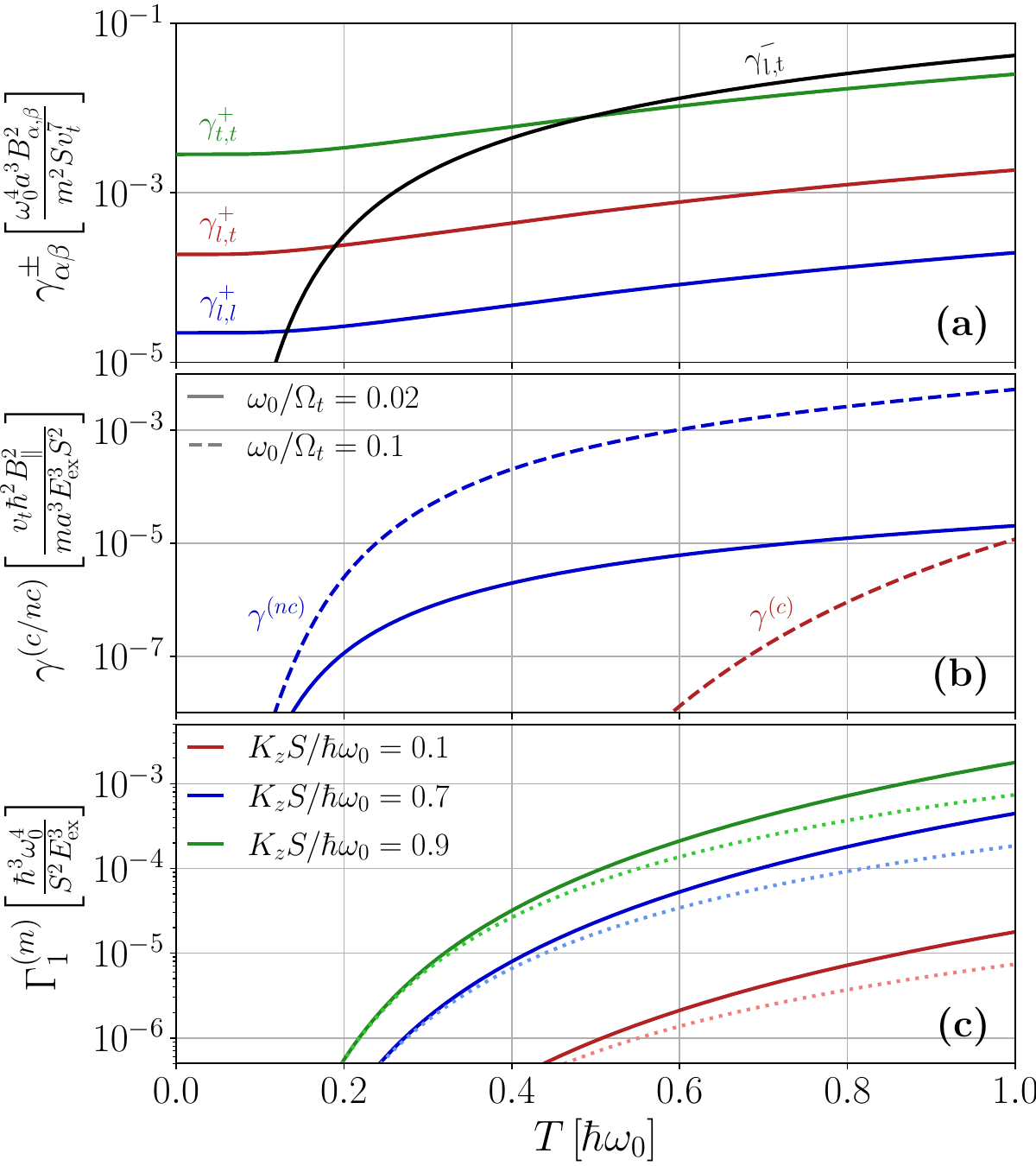}
    \caption{The different contributions to the relaxation rate of a magnon qubit at $\bk_0=0$, from (a) one-magnon--two-phonon processes, (b) two-magnon--one-phonon processes, and (c) four-magnon processes where the dotted lines show the approximate analytical low-temperature result from Eq.~\eqref{eq:MagnonDecay}. We have set $v_l=2v_t$ and in (b) we have set $B_\perp=2B_\parallel$, as appropriate for YIG \cite{Streib}.}
    \label{fig:PhonT1}
\end{figure}

In Fig.~\ref{fig:PhonT1} we plot the contributions to the relaxation rate derived above for the uniform mode $\bk_0=0$, evaluated in the weak-damping limit $\eta_\bk, \eta_{\bq,\lambda} \to 0$, where the Lorentzian line shapes in Eq.~\eqref{eq:FullGammas} are treated as delta functions. We set $v_l=2v_t$, a relation that holds approximately for many materials, including YIG~\cite{Streib}.
Figure~\ref{fig:PhonT1}(a) shows the four distinct one-magnon--two-phonon contributions.
Figure~\ref{fig:PhonT1}(b) shows the contribution from two-magnon--one-phonon processes for two different ratios of $\omega_0/\Omega_t$. Here $\gamma^{(\text{nc})}\gg\gamma^{(\text{c})}$ since, for realistic qubit frequencies $\omega_0 \ll \Omega_{l,t}$, magnon-conserving scattering involves only high-frequency magnons and phonons which are strongly suppressed at low temperatures but become important at high temperatures $T/\hbar\omega_0\gg1$ \cite{Streib}.
Finally, Fig.~\ref{fig:PhonT1}(c) presents the relaxation rate due to magnon-magnon interactions for three ratios of $K_zS/\hbar\omega_0$.

As expected, only the contributions $\gamma^+_{\alpha,\beta}$ from one-magnon--two-phonon processes [Fig.~\ref{fig:PhonT1}(a)] survive in the limit $T\to 0$, corresponding to the decay of the qubit excited state into two phonons. In contrast, the decay of a qubit excitation into a magnon and a phonon is forbidden by energy conservation, since the chosen qubit mode $\bk_0=0$ lies at the bottom of the magnon dispersion.
If finite damping $\eta_\bk, \tilde{\eta}_{\bq,\lambda} > 0$ were included, the broadened Lorentzian line shapes would yield a finite contribution to $\Gamma^{(21)}_1$ from off-resonant processes even at zero temperature. However, because of the off-resonant nature of this contribution, it would require the introduction of a momentum cutoff to ensure convergence, and it thus cannot be treated reliably within the continuum approximation.
Furthermore, in geometries where dipolar interactions produce minima in the magnon dispersion at finite wave vectors $\bk$~\cite{YIG_FilmDisp}, the decay of a $\bk_0=0$ magnon into a magnon and a phonon becomes allowed.

We now analyze the prefactors of the different contributions for the prototypical case of YIG.
For YIG at room temperature the lattice constant $a=12.376\,\text{Å}$, exchange energy $E_\text{ex}=3.45\,\text{meV}$  and sound velocities are $v_{t}=3843\,\text{m/s}$ and $v_{l}=7209\,\text{m/s}$ \cite{Streib}.
The various magnon--phonon coupling constants have been determined experimentally to be $B_\parallel=4.12\,\text{meV}$, $B_\perp=8.24\,\text{meV}$, $B_{144}=-6\,\text{meV}$, $B_{155}=-44\,\text{meV}$, and $B_{456}=-32\,\text{meV}$ (see Appendix~\ref{secA:Hams} for the notation).
For a qubit frequency $\omega_0/2\pi = 20\,\text{GHz}$, we find the prefactors $\omega_0^4 a^3 B^2_{t,t} / m^2 S v_t^7 \approx 8 \times 10^{-4}\,\text{s}^{-1}$ and $B^2_{l,t}/B^2_{t,t}\approx3$ for the one-magnon--two-phonon processes, $v_t \hbar^2 B_\parallel^2 / (m a^3 E_\text{ex}^3 S^2) \approx 3 \times 10^4\,\text{s}^{-1}$ for the two-magnon--one-phonon processes, and $\hbar^3\omega_0^4 / S^2 E_\text{ex}^3 \approx 9 \times 10^3\,\text{s}^{-1}$ for the four-magnon processes.

The strong suppression of the two-phonon processes arises from two main factors: First, the heavy YIG unit cell results in a minimal characteristic strain per phonon excitation, $X = \sqrt{\hbar / m v_{l,t} a} \sim 10^{-3}\ll 1/\sqrt{S}$, leading to much weaker coupling to the phonons compared to the magnons. Second, because of the small qubit frequency $\omega_0 \ll \Omega_{l,t}$, spontaneous decay processes necessarily involve only low-frequency environmental modes, while scattering processes which preserve the number of quasiparticles can involve high-frequency modes as well. This leads to smaller wave vectors for spontaneous processes than for scattering processes, e.g., the wave vector for spontaneous two-phonon processes $ \sim \omega_0 / v_{t,l}$ is much smaller than the one for two-magnon scattering processes $ \sim \Omega_{l,t} / v_{t,l}$. This reduces not only the phase space available for scattering $\propto (aq)^2$, but also the magnon--phonon coupling constants [see Eqs.~\eqref{eqA:1PhonCoupl}, \eqref{eqA:2PhonCoupl} and \eqref{eq:ExchCouplPhon}], which come with a factor of the effective strain $X\sqrt{qa}$ per phonon mode. The latter argument also applies to the $s = -$ branch of the magnon-nonconserving two-magnon processes, where the relevant wave vector $\sim \omega_0 / v_{t,l}$ is likewise suppressed. Similarly, even if allowed by dipolar interactions, the relaxation of the $\bk_0=0$ magnon decaying into a magnon and phonon would also be permitted only for small wave vectors \cite{YIG_FilmDisp} and thus would be small compared to the thermal rates at room temperature. However, due to a weaker dependence on the wave vector $q$ and mass $m$, such a decay process could still be stronger than the two-phonon process, depending on the shape and geometry of the sample.

We note that in current millikelvin experiments~\cite{YIGMilliKelvin1,YIGMilliKelvin2}, the observed low-temperature damping for the uniform mode is predominantly attributed to coupling to two-level fluctuators, which is believed to dominate over the intrinsic relaxation processes considered here. Since these two-level systems are associated with structural imperfections like impurities and surface roughness, the rates calculated in this work provide a fundamental lower bound on the total qubit relaxation rates. The same is likely true for magnons at finite wave vectors~\cite{AndriiDamping}.
The relatively small contributions to relaxation from magnon-magnon and magnon-phonon coupling we find here are consistent with these observations.

\subsection{Pure dephasing}\label{sec:deph}

We next turn to the pure dephasing rate $\Gamma_\phi$.
To leading order, the only relevant interaction terms read
\begin{subequations}
    \begin{align}
    \hat v_{11}^{(m)}  &= \sum_{\bk \neq \bk_0}\tilde{B}^{(m)}_{\bk}
    \big( \hb_{\bk}^\dagger\hb_{\bk}
    - \langle \hb_{\bk}^\dagger\hb_{\bk} \rangle \big),\\
    \hat v_{00}^{(m)}  &= 0,
    \end{align}
\end{subequations}
which straightforwardly yield
\begin{equation}\label{eq:gammaphi}
    \Gamma^{(m)}_\phi = \int \!\! \frac{\td t}{4\hbar^2}  \sum_{\bk \neq \bk_0}\big[ \tilde{B}^{(m)}_{\bk} \big]^2 \big\langle \{ \hat{\tilde n}_{\bk} (t) , \hat{\tilde n}_\bk (0) \}\big \rangle,
\end{equation}
where we introduced $\hat{\tilde n}_{\bk}(t) = \hb_{\bk}^\dagger\hb_{\bk}(t) - \langle \hb_{\bk}^\dagger\hb_{\bk} \rangle $ to denote the fluctuations of the occupation number operator~\footnote{Since $\hat{\tilde{n}}_{\bk}(t)$ commutes with the environmental Hamiltonian $\hat{H}_E = \sum_{\bk\neq\bk_0}\hbar\omega_\bk\hb^\dagger_{\bk}\hb_{\bk}^{\phantom{\dagger}}$, the integrand in Eq.~\eqref{eq:gammaphi} becomes time-independent in the absence of damping. As a result, the integral, and thus the dephasing rate, formally diverges.
This divergence originates from the Born-Markov approximation employed in deriving the Nakajima-Zwanzig equation, where it was assumed that the bath's coherence time is short compared to the system's. Without damping, the evolution is governed solely by $\hat{H}_E$, resulting in an infinite bath coherence time and invalidating the Born-Markov approximation.
Previous works addressed this issue by introducing a relaxation of momentum conservation among the bath modes~\cite{YuanDephasing,MagnonQubit}, for instance, due to impurities or boundary effects.
Here, we instead exploit the inclusion of finite damping in the environmental magnon Green's functions, Eq.~\eqref{eq:KeldyshGreenMag}, parameterized by $\eta_\bk$, to regularize the integral without introducing additional assumptions.}.

Using the Keldysh expectation value from Eq.~\eqref{eq:KeldyshExp}, we then find
\begin{align}
        \Gamma^{(m)}_\phi = \int \! \frac{\td \omega}{4\pi \hbar^2}
        \sum_{\bk \neq \bk_0}
        \frac{ \big( \tilde{B}^{(m)}_{\bk} \big)^2}{
        \sinh^{2}(\hbar\omega/2T )}
        \bigg[\frac{\eta_\bk}{(\omega-\omega_\bk)^2+\eta_\bk^2}\bigg]^2.
\label{eq:DeltaNk}
\end{align}
We assume that Gilbert damping $\eta_\bk=\alpha\omega$ \cite{TakeiKeldysh} describes the damping rate well. 
If the magnon modes are weakly damped, $\alpha\ll 1$, then the integrand in Eq.~\eqref{eq:DeltaNk} is sharply localized around $\omega_\bk$, and we can approximate the integral by
\begin{align}\label{eq:TphiS}
        \Gamma^{(m)}_\phi \approx\frac{1}{\hbar^2} 
        \sum_{\bk \neq \bk_0}
        \frac{ \big( \tilde{B}^{(m)}_{\bk} \big)^2}{ 8 \alpha \omega_\bk
        \sinh^{2}(\hbar\omega_\bk /2T )}.
\end{align}
The fact that this quantity diverges for $\alpha\to 0$ reflects that the Born-Markov approximation fails when the coherence time of the bath exceeds that of the system.
Consequently, this expression for the dephasing rate is formally valid only in the regime $\Gamma_\phi \ll \eta_\bk$, where the Markovian approximation remains justified.

We again convert the sum in Eq.~\eqref{eq:TphiS} into an integral and assume the same quadratic dispersion given in Eq.~\eqref{eq:disp} in the long wavelength limit $ka \ll 1$.
Again for $\bk_0 = 0$ we can then arrive at a closed-form expression in the low-temperature limit $\hbar \omega_0 \ll T$,
\begin{equation}\label{eq:dephK}
    \Gamma_\phi \approx \frac{1}{4\pi^{3/2}N} \frac{K_z^2\sqrt{\hbar\omega_0}}{ \hbar\alpha E_\text{ex}^{3/2}}\bigg(\frac{T}{\hbar\omega_0}\bigg)^{3/2}e^{-\hbar\omega_{0}/T},
\end{equation}
where the contribution from the exchange interaction in the coupling energies $\tilde B_{\bk}^{(m)}$ vanishes at $\bk_0=0$ [see Eq.~(\ref{eq:bkex})]. But as the exchange and anisotropy contributions to the coupling have opposite symmetries under inversion $\bk\to-\bk$, they give separable contributions $\Gamma_\phi = \gamma_K+\gamma_\text{ex}$, and we can extend our analytical expression to qubit modes where $\bk_0 \neq 0$. If we still assume the quadratic dispersion from Eq.~\eqref{eq:disp}, in the low-temperature limit $\gamma_K$ is as given in Eq.~(\ref{eq:dephK}), and
\begin{align}\label{eq:gammaexLT}
        \gamma_\text{ex} \approx \frac{a^2|\bk_0|^2}{8\pi^{3/2}N}\frac{(\hbar\omega_0)^{3/2}}{ \hbar S^2\alpha\sqrt{E_\text{ex}}}\bigg(\frac{T}{\hbar\omega_0}\bigg)^{5/2} e^{-\hbar\omega_{\bk_0}/T}.
\end{align}

The rates $\gamma_K$ and $\gamma_\text{ex}$ scale with both $\alpha^{-1}$ and $N^{-1}$, and thus differ markedly from the relaxation processes described in Eq.~\eqref{eq:MagnonDecay} and dephasing rates when momentum conservation is relaxed \cite{YuanDephasing}, which are independent of both. The increased dephasing rate for smaller Gilbert damping $\alpha$ follows because an increased coherence time of the environment leads to a larger cumulative uncertainty in the qubit splitting. At the same time, the $N$ dependence is a consequence of the reduced available phase space. The temperature-scaling of $\gamma_K$ and $\gamma_\text{ex}$ are also due to the differing $k$ dependence of the coupling constants $\big[\tilde{B}^{(m)}_{\bk}\big]^2$.

\begin{figure}[t!]
    \centering
    \includegraphics[width=\linewidth]{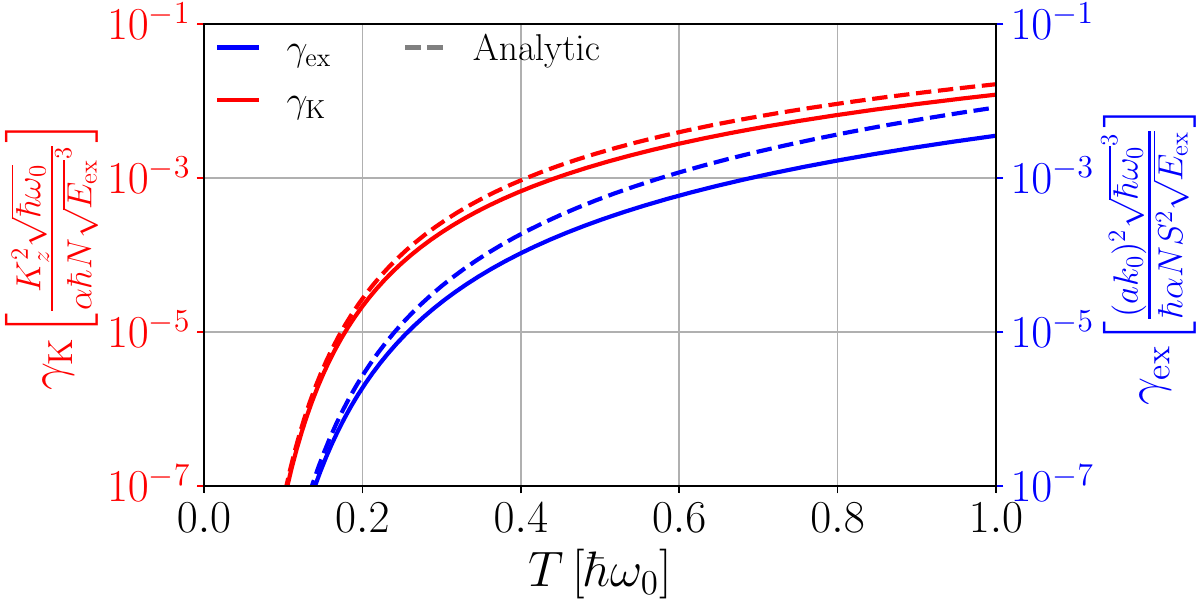}
    \caption{The contribution to the dephasing rate from anisotropy- (red line) and exchange-mediated (blue line) interaction. The dashed lines show the corresponding low-temperature analytical solutions from Eqs.~\eqref{eq:dephK} and \eqref{eq:gammaexLT}, respectively.}
    \label{fig:TPhi}
\end{figure}

In Fig.~\ref{fig:TPhi} we plot the contributions $\gamma_{K}$ and $\gamma_\text{ex}$ as a function of $T/\hbar \omega_{0}$, both for the numerical solution of the integral limit of Eq.~\eqref{eq:TphiS} (solid lines) and for the low-temperature analytic expressions in Eqs.~\eqref{eq:dephK} and \eqref{eq:gammaexLT} (dashed lines). For YIG, again using room-temperature parameters and $\omega_0/2\pi=20\,\text{GHz}$, we find the prefactors $K_z^2\sqrt{\hbar\omega_0}/ \hbar \sqrt{E_\text{ex}}^{3}\approx 2 \times 10^6 (K_z S/\hbar\omega_0)^2\,\text{s}^{-1}$ and $\sqrt{\hbar\omega_0}^{3}/\hbar S^2\sqrt{E_\text{ex}}\approx 10^8\,\text{s}^{-1}$. Thus, the dephasing rate can be sizable for small and clean magnets. Furthermore, for $\alpha N \lesssim 1$ the Born--Markov approximation $\Gamma_\phi/(\alpha\omega_0)\ll 1$ is satisfied.

\section{\label{sec:Concs}Conclusion}

We calculated the decoherence rates for a magnon-based qubit in a ferromagnetic insulator due to magnon-phonon and magnon-magnon interactions. We further focused on the case where the magnon consists of a uniform Kittel mode, giving an analytical result for a quadratic magnon dispersion in the long-wavelength limit. We found that the only allowed zero-temperature relaxation process, where the qubit relaxes by emitting two phonons, is heavily suppressed in YIG due to the heavy unit cell and the low frequency of the scattered phonons. In geometries where dipolar interactions lead to minima in the dispersion at finite wave vectors, the relaxation rate from the qubit decaying into a magnon and phonon could thus be dominant. Still, it would also be expected to be similarly suppressed. This agrees well with low-temperature experiments, where the dominant damping process was attributed to surface imperfections and impurities \cite{YIGMilliKelvin1,YIGMilliKelvin2, AndriiDamping}. By going beyond the continuum approximation, future work could also use our theory to calculate off-resonant contributions, incorporating the effects of finite broadening of the environment.

We also showed how the dephasing rate scales inversely with the size and damping rate of the magnet when enforcing momentum conservation. This expression is, however, valid only if the dephasing rate is slower than Gilbert damping; the opposite signals the onset of non-Markovian dynamics, which cannot be treated under the Born-Markov approximation. Such a regime would also be sensitive to the qubit state preparation. We showed that this becomes especially important for small and clean magnets, which are attractive for quantum devices. A future work could try to bridge the gap, by taking care of the fully retarded nature of the interactions.

\section*{Acknowledgements}
We thank A. Qaiumzadeh for useful discussions.
The Research Council of Norway has supported this project through its Centres of Excellence funding scheme, Project No. 262633, ``QuSpin''. 

\appendix

\section{\label{secA:Hams}Interaction vertices}

\emph{Magnon--magnon interaction Hamiltonian---}We consider an easy-axis ferromagnet with ferromagnetic exchange interactions, subject to an external magnetic field $B$ along the easy axis. We describe the system with the Hamiltonian \cite{YIG_FilmDisp, YIGDispAndInteractions}
\begin{align}
    \begin{split}
            H_\text{mag}=&-\frac{1}{2}\sum_{i,j}J_{ij}\bS_i\cdot \bS_j-\frac{K_z}{2}\sum_i(S_i^z)^2\\
            &-g\mu_\text{B} B\sum_i S_i^z,
    \end{split}\label{eqA:FMHamRealSpace}
\end{align}
where $i$, $j$ label the lattice sites at positions $\bR_{i}$ hosting spins $\bS_{i}$, $J_{ij}$ is the exchange-interaction strength between spins at sites $i$ and $j$, $K_z$ is the strength of the easy axis, the $g$ factor $g\approx2$, and $\mu_B$ is the Bohr magneton.

To proceed, we express the spin operators in terms of bosonic magnon operators via the Holstein–Primakoff transformation:  $S^z_i=S-\hb^\dagger_i\hb_i$, $S^+_i=\sqrt{2S}(1-\hb^\dagger_i\hb_i/4S)\hb_i$, and $S^-_i=\sqrt{2S}\hb^\dagger_i(1-\hb^\dagger_i\hb_i/4S)$. Here, the magnon creation and annihilation operators $\hb^{(\dagger)}$ satisfy bosonic commutation relations, and $S^{\pm}=S^x\pm iS^y$. To second order in the operators $\hb^{(\dagger)}$, the Hamiltonian \eqref{eqA:FMHamRealSpace} is diagonalized by using the Fourier transform $\hb_i=N^{-1/2}\sum_\bk e^{i\bk\cdot\bR_i}\hb_\bk$, where $N$ is the number of unit cells, which gives
\begin{subequations}
    \begin{align}
        \hH_\text{mag}&=\hbar\sum_\bk \omega_\bk\hb_\bk^\dagger\hb_\bk,\\
        \hbar\omega_\bk=&S(J_0-J_\bk)+SK_z +g\mu_\text{B}B.
    \end{align}
\end{subequations}
Here, $J_\bk=\sum_{j \in i \text{'s n.}}J_{ij}\cos[ \bk\cdot (\boldsymbol{R}_j - \boldsymbol{R}_i) ]$, where $j$ is summed over all sites neighboring site $i$ and the lattice has been assumed inversion symmetric. For a cubic lattice with only nearest neighbor interactions, we find in the long-wavelength limit $ka\ll1$ that $S(J_0-J_\bk)\approx E_\text{ex}k^2a^2$, where $a$ is the lattice constant, $E_\text{ex}=2JS$, and $J$ is the nearest neighbor hopping strength. The level splitting between the $k$ modes is $\sim 4\pi^2 E_\text{ex} N^{-2/3}$.
For YIG at room temperature $E_\text{ex} \approx 3.45\,\text{meV}$, so having a gap of at least $1\,\text{GHz}$ requires $\lesssim 10^7$ YIG unit cells.

Expanding $H_\text{mag}$ to fourth order in the operators $\hb$ and $\hb^\dagger$ gives the additional term~\cite{TakeiKeldysh}
\begin{align}
     \hH_4=&-\frac{1}{2N}\sum_{\{\bk_i\}}B_{\bk_1,\bk_3}\delta_{\bk_1+\bk_2,\bk_3+\bk_4}\hb^\dagger_{\bk_1}\hb^\dagger_{\bk_2}\hb_{\bk_3}\hb_{\bk_4},
\end{align}
where
\begin{align}
        B_{\bk,\bk'}=&K_z+J_{\bk-\bk'}-\frac{1}{2}(J_{\bk}+J_{\bk'}).
\end{align}
By focusing on interactions including a specific magnon mode $\bk_0$, we recover the Hamiltonian \eqref{eq:Mag4Int} given in the main text, where the symmetrized interaction strengths are given by
\begin{subequations}
\begin{align}
\begin{split}
    B^{(m)}_{\bk,\bk'}&= -\frac{B_{\bk_0,\bk'}+B_{\bk,\bk'}+B_{\bk_0,\bk_0+\bk-\bk'}+B_{\bk,\bk_0+\bk-\bk'}}{4N},\end{split}\\
    \tilde{B}^{(m)}_{\bk}&= -\frac{B_{\bk_0,\bk_0}+B_{\bk,\bk}+2B_{\bk_0,\bk}}{2N}.\label{eq:bkexF}
    \end{align}
\end{subequations}
In the long-wavelength limit, we find
\begin{subequations}
\begin{align}
\begin{split}
    B^{(m)}_{\bk,\bk'}&= -\frac{1}{2NS}\Big\{2K_zS + E_{\rm ex}a^2 \big[\bk_0\cdot\bk\\
    &\hspace{1cm}-\bk'\cdot(\bk'-\bk_0-\bk) \big] \Big\},\end{split}\\
    \tilde{B}^{(m)}_{\bk}&= -\frac{2}{NS}\big[ K_zS + E_{\rm ex} a^2 (\bk\cdot\bk_0) \big].\label{eq:bkex}
    \end{align}
\end{subequations}

\emph{Magnon--Phonon interaction Hamiltonian---}We describe the phonon bath with the Hamiltonian
\begin{equation}
    H_\text{ph}=\hbar\sum_{\bq,\lambda}\tilde{\omega}_{\bq,\lambda}\ha_\bq^\dagger\ha_\bq.
\end{equation}
where $\ha_\bq$ ($\ha^\dagger_\bq$) destroys (creates) a phonon with polarization $\lambda$ at wave vector $\bq$ with frequency $\tilde \omega_{\bq,\lambda}=v_\lambda q$. In this work, we exclusively consider low-energy acoustic phonons with a linear dispersion.
We assume the long-wavelength limit $q a \ll 1$, and for the two transverse phonon modes we write $v_1=v_2=v_t$ and for the longitudinal mode $v_3=v_l$.

Phonons are the quantized units of lattice vibrations, and the bosonic operators $\ha^\dagger$ can be connected to the displacement $\boldsymbol{X}_i=\boldsymbol{r}_i-\boldsymbol{R}_i$ of the atoms labeled $i$ away from their equilibrium positions $\boldsymbol{R}_i$ via
\cite{Streib, YIGDispAndInteractions}
\begin{equation}
    \boldsymbol{X}_i=\frac{1}{\sqrt{N}}\sum_{\bq,\lambda}\be_{\bq,\lambda}
    \sqrt \frac{\hbar}{2m\tilde{\omega}_{\bq\lambda}}\big(\ha_{\bq\lambda}+\ha^\dagger_{-\bq\lambda}\big) e^{i\bq\cdot\boldsymbol{R}_i}.
\end{equation}
Here, $\bq$ is the phonon wave vector and $N$ denotes the number of lattice sites.
In terms of the spherical angles $\{\theta_\bq,\phi_\bq\}$ of the vector $\bq$, the (complex) unit vectors $\be_{\bq,\lambda}$ characterizing the three phonon polarizations can be written as
\begin{subequations}
    \begin{align}
        \be_{\bq,1}&=(\cos\theta_\bq\cos\phi_\bq,\cos\theta_\bq\sin\phi_\bq,-\sin\theta_\bq),\\
        \be_{\bq,2}&=i(-\sin\phi_\bq,\cos\phi_\bq,0),\\
        \be_{\bq,3}&=i(\sin\theta_\bq\cos\phi_\bq,\sin\theta_\bq\sin\phi_\bq,\cos\theta_\bq),
    \end{align}
\end{subequations}
where $\lambda=1,2$ correspond to transverse modes and $\lambda=3$ corresponds to the longitudinal mode, as before.
In the following, we express the couplings in terms of the dimensionless displacement operators $\hat{\xi}_{\bq\lambda}= \ha_{\bq\lambda}+\ha^\dagger_{-\bq\lambda}$.

Following Streib \textit{et al.}~\cite{Streib}, we now turn to the interaction between phonons and magnons, considering only terms to third order in the magnon and phonon operators. We include two distinct mechanisms that contribute to magnon--phonon coupling: (1) coupling through magnetoelastic anisotropy, i.e., the emergence of additional anisotropic spin--spin interactions due to the breaking of crystal symmetries by the lattice deformations, and (2) coupling via local modulations of the nearest-neighbor exchange interaction due to the lattice deformations.

(1) The anisotropy contribution to the magnetoelastic energy is, to lowest order, described by the Hamiltonian~\cite{AniEnergy1,Streib,YIGDispAndInteractions}
\begin{align}
    	\begin{split}
    		H_{\text{me}}^{(1)}=&\frac{1}{S^2}\sum_i\sum_{\alpha,\beta \in \{x,y,z\} } B_{\alpha\beta} S_i^\alpha S_i^\beta X_i^{\alpha\beta},
    	\end{split}
\end{align}
where $X_i^{\alpha\beta}=(\partial X_i^\alpha/\partial R_i^\beta+\partial X_i^\beta/\partial R_i^\alpha)/2$ is the symmetric strain tensor.
For a lattice with cubic symmetry, the contributions to this energy are restricted to the terms $B_{\alpha\beta}=B_\parallel\delta_{\alpha,\beta}+(1-\delta_{\alpha,\beta})B_\perp$.
Using the Holstein--Primakoff transformation again, we find to leading order
\begin{widetext}
\begin{align}
\hat{H}_{\text{me}}^{(1)} = \sum_{\bk,\lambda} (B_{\bk,\lambda}^{\bar{b}}\hb_{\bk}^\dagger\hat{\xi}_{\bk\lambda}+\text{H.c}) + \frac{1}{\sqrt{N}} \sum_{\bk,\bk'} \sum_{\bq,\lambda} \hat{\xi}_{\bq\lambda} \bigg[\delta_{\bk,\bk'+\bq} B^{\bar{b}b}_{\bq\lambda} \hb^\dagger_\bk\hb_{\bk'} 
 + \delta_{\bk+\bk',-\bq}\frac{B^{bb}_{\bq\lambda}}{2} \hb_\bk\hb_{\bk'}+\delta_{\bk+\bk',\bq}\frac{B^{\bar{b}\bar{b}}_{\bq\lambda}}{2}\hb_\bk^\dagger\hb_{\bk'}^\dagger\bigg],
\end{align}
where the interaction vertices read
\begin{subequations}
    \begin{align}
	B_{\bk,\lambda}^{\bar{b}} =& \frac{B_\perp}{\sqrt{2S}}\sqrt{\frac{\hbar}{2m\tilde{\omega}_{\bq,\lambda}}}\Big[iq_ze_{\bq,\lambda}^x-q_ze_{\bq,\lambda}^y +(iq_x-q_y)e_{\bq,\lambda}^z\Big],\\
    B^{\bar{b}b}_{\bq,\lambda}=&\frac{iB_\parallel}{S}\sqrt{\frac{\hbar}{2m\tilde{\omega}_{\bq,\lambda}}}\Big(q^xe_{\bq,\lambda}^x+q^ye_{\bq,\lambda}^y-2q^ze_{\bq,\lambda}^z\Big),\\
    \begin{split}
    B^{\bar{b}\bar{b}}_{\bq,\lambda}=&\frac{iB_\parallel}{S}\sqrt{\frac{\hbar}{2m\tilde{\omega}_{\bq,\lambda}}}\Big(q^xe_{\bq,\lambda}^x-q^ye_{\bq,\lambda}^y\Big) -\frac{B_\perp}{S}\sqrt{\frac{\hbar}{2m\tilde{\omega}_{\bq,\lambda}}}\Big(q^xe_{\bq,\lambda}^y+q^ye_{\bq,\lambda}^x\Big),
    \end{split}
\end{align}\label{eqA:1PhonCoupl}
\end{subequations}
and $B^{\bar{b}\bar{b}}_{\bq\lambda}=(B^{bb}_{-\bq\lambda})^*$.

There are also relevant terms in the magnetoelastic energy that are second order in the displacement but first order in magnon operators~\cite{Eastman},
\begin{align}
        H_{\text{me}}^{(2)}=&\frac{1}{S^2}\sum_{i} \sum_{\alpha,\beta,\gamma \in P_{xyz}} \big(2B_{144} S_i^\alpha S_i^\beta X_i^{\alpha\beta} X_i^{\gamma\gamma} + 4B_{155}S_i^\alpha S_i^\beta X_i^{\alpha\beta} X_i^{\alpha\alpha} + 8B_{456} S_i^\alpha S_i^\beta X_i^{\alpha\gamma} X_i^{\beta\gamma}\big),
\end{align}
where $P_{xyz}$ denotes the set of all permutations of the three coordinates $\{x,y,z\}$, i.e., the sum goes over all $\alpha\neq\beta\neq\gamma$.
We have neglected the terms $\propto B_{111}$, $B_{123}$, and $B_{441}$, which permit only contributions that are $\propto S^\alpha S^\alpha$ and thus are even in the operators $\hb^{(\dagger)}$ and fourth order in the magnon--phonon coupling.
In terms of magnon and phonon operators, this yields to leading order~\cite{Streib}
\begin{align}
        \hat{H}_{\rm me}^{(2)} = -\frac{1}{\sqrt{N}}&\sum_{\bk,\bq_1,\bq_2}\sum_{\lambda_1,\lambda_2} B_{\bq_1\lambda_1,\bq_2\lambda_2} \delta_{\bk,\bq_1+\bq_2}\hb_\bk^\dagger\hX_{\bq_1\lambda_1}\hX_{\bq_2\lambda_2} + \text{H.c.},
\end{align}
with the symmetrized interaction vertex
\begin{align}\label{eqA:2PhonCoupl}
B_{\bq_1\lambda_1,\bq_2\lambda_2}=&\tilde B_{\bq_1\lambda_1,\bq_2\lambda_2}+\tilde B_{\bq_2\lambda_2,\bq_1\lambda_1},\\
\begin{split}
    \tilde{B}_{\bq_1\lambda_1,\bq_2\lambda_2}=&\frac{\hbar}{2m\sqrt{\tilde{\omega}_{\bq_1\lambda_1}\tilde{\omega}_{\bq_2\lambda_2}}}\sqrt{\frac{2}{S}} \big\{ {}  {} B_{144}[ i I_1^{xx}I_2^{yz} + I_1^{yy}I_2^{xz}] + B_{155}[ i (I_1^{yy}+I_1^{zz})I_2^{yz}+  (I_1^{xx}+I_1^{zz})I_2^{xz}] \\
    {} & {}\hspace{3.5cm} + 2 B_{456}[iI_1^{xz}I_2^{xy}+ I_1^{yz}I_2^{xy}]\big\},
\end{split}
\end{align}%
where $I_n^{\alpha\beta} = (e_{\bq_n\lambda_n}^\alpha q_n^\beta + e_{\bq_n\lambda_n}^\beta q_n^\alpha)/2$.

(2) The contribution from the exchange coupling follows from considering the effect of lattice deformations on the nearest-neighbor exchange interaction.
Assuming an isotropic nearest-neighbor exchange $J_{ij} = J(r_{ij})$, with $r_{ij}$ being the distance between spins $i$ and $j$, one can expand for small displacements $J_{ij} \approx J(a) + (\td J/\td r) (\lvert\br_i-\br_j\rvert-a)$.
For a cubic lattice, this yields the magnetoelastic energy
\begin{align}\label{eq:exchange}
        {H}_{\rm ex} = -\frac{1}{2}\frac{\td J}{\td r} \sum_{\boldsymbol n} \sum_{\alpha\in \{x,y,z\} } \big[ \big (X_{\boldsymbol{n} + \hat{\alpha} }^\alpha - X_{\boldsymbol n}^\alpha\big)\bS_{\boldsymbol{n} + \hat{\alpha} } + \big (X_{\boldsymbol{n} }^\alpha - X_{\boldsymbol n - \hat{\alpha}}^\alpha\big)\bS_{\boldsymbol{n} - \hat{\alpha} } \big]\cdot\bS_{\boldsymbol n},
\end{align}
where now, instead of labeling the spins $i$, we use the lattice vectors $\boldsymbol n = (n_x, n_y, n_z)$ where $n_{x,y,z}$ are integers, such that $\boldsymbol R_i = \boldsymbol n\, a$.
The vector $\hat \alpha$ indicates the unit vector along the $\alpha$-direction.
Expressed in terms of magnon and phonon operators, we find
\begin{align}
    \hat H_{\rm ex} =\frac{1}{\sqrt{N}}\sum_{\bk,\bk'}\sum_{\bq,\lambda}B_{\bk,\bq\lambda}^\text{ex}\delta_{\bk,\bk'+\bq}\hb^\dagger_\bk\hb_{\bk'}\hat{\xi}_{\bq\lambda}.
\end{align}
In the long-wavelength limit, we find for the interaction vertex
\begin{equation}
    B_{\bk,\bq\lambda}^\text{ex}=i  a^3 S\frac{\td J}{\td r} \sqrt{\frac{\hbar}{2m\tilde{\omega}_{\bq,\lambda}}}\sum_\alpha e^\alpha_{\bq,\lambda}k^\alpha q^\alpha(k^\alpha-q^\alpha).\label{eq:ExchCouplPhon}
\end{equation}

The parameters in the magnon--phonon interaction terms in Eq.~\eqref{eq:MagPhonInteraction} can thus be connected to the vertices derived above as follows,
\begin{subequations}
    \begin{align}
    B^{(11)}_{\lambda}&=B^{\bar{b}}_{\bk_0,\lambda},\\
    B^{(12)}_{\bq,\lambda_1,\lambda_2}&=\frac{-1}{\sqrt{N}}B_{\bq\lambda_1,(\bk_0-\bq)\lambda_2},\\
    B^{(21)}_{\bq,\lambda}&=\frac{1}{\sqrt{N}}\big(B^{\bar{b}b}_{\bq,\lambda}+B^{\text{ex}}_{\bk_0,\bq\lambda}\big),\\
    \tilde{B}^{(21)}_{\bq,\lambda}&=\frac{1}{\sqrt{N}}B^{\bar{b}\bar{b}}_{\bq,\lambda}.
    \end{align}
\end{subequations}
Furthermore, in Eq.~\eqref{eq:2PhonRates} we use the following short-hand notation for the effective two--phonon coupling constants,
\begin{subequations}
    \begin{align}
            B_{t,t}^2&=\frac{1}{2}\frac{20(B_{144}-B_{155})^2-4B_{456}(B_{144}-B_{155})+59B_{456}^2}{45},\\
            B_{l,t}^2&=\frac{7(B_{144}+2B_{456})^2+4B_{155}(B_{144}+2B_{456}+13B_{155})}{45},\\
            B_{l,l}^2&=4\frac{(B_{144}+2B_{456})^2+4B_{155}(B_{144}+2B_{456}+4B_{155})}{45}.
    \end{align}\label{eqA:BeffTwoPhonon}
\end{subequations}

\section{Full expressions for the relaxation rates}
\label{app:explicit}

The full expressions for the four contributions to the relaxation rate $\Gamma_1$, including the effect of the finite lifetime of the environmental magnons, read
\begin{subequations}\label{eq:relaxationfiniteeps}
\begin{align}
\Gamma_{1}^{(12)} = {} & {} \frac{1}{\hbar^2} \!\! \sum_{\bq,\lambda_1,\lambda_2} \!\! \big|B^{(12)}_{\bq,\lambda_1,\lambda_2}\big|^2\int\td t\,e^{i\omega_{\bk_0} t}\iint
\td \omega_1\td\omega_2\,L(\omega_1-\tilde z_{\bq,\lambda_1}){L}(\omega_2-\tilde z_{\bq-\bk_0,\lambda_2})\nonumber \\
 {} & {} \quad \times\bigg\{N^+(\omega_1,\omega_2)\Big[e^{i(\omega_1+\omega_2)t}+e^{-i(\omega_1+\omega_2)t}\Big]+2N^-(\omega_1,\omega_2)e^{i(\omega_1-\omega_2)t}\bigg\},\\
 \Gamma_{1}^{(21)} = {} & {} \frac{1}{2\hbar^2} \sum_{\bq,\lambda}  \int\td t\,e^{i\omega_{\bk_0} t}\iint
\td \omega_1\td\omega_2\,L(\omega_1-z_\bq){L}(\omega_2-\tilde z_{\bq-\bk_0,\lambda})
\nonumber \\ {} & {} \quad \times 
\sum_{\sigma = \pm} N^\sigma(\omega_1,\omega_2)\Big[\big|B^{(21)}_{\bq,\lambda}\big|^2e^{-i(\omega_1+\sigma\omega_2)t}+\big|\tilde{B}^{(21)}_{\bq,\lambda}\big|^2e^{i(\omega_1+\sigma\omega_2)t}\Big],\\
 \Gamma_{1}^{(m)} = {} & {} \frac{2}{\hbar^2} \sum_{\bk,\bk'}\big|B^{(m)}_{\bk,\bk'}\big|^2\int\td t\,e^{i\omega_{\bk_0} t}\iiint \td\omega_1\td\omega_2\td\omega_3\,L(\omega_1-z_\bk)L(\omega_2-z_{\bk'})L(\omega_3-z_{\bk_0+\bk-\bk'}) \nonumber\\
 {} & {} \quad\times \big\{ [n(\omega_1)+1] n(\omega_2)n(\omega_3) + n(\omega_1)[n(\omega_2)+1][n(\omega_3)+1]\big\} e^{i(\omega_1-\omega_2-\omega_3)t},
\end{align}
\end{subequations}
\end{widetext}
where $L(z)$ is the Lorentzian from Eq.~\eqref{eq:Lorentzian}.

These integrals do not have closed-form solutions. However, when $\eta_\bk,\tilde \eta_{\bq,\lambda} \ll \omega_\bk, \tilde{\omega}_{\bq,\lambda}$ all Lorentzians are sharply peaked, suggesting that in that case only frequencies within a few $\eta_\bk, \tilde \eta_{\bq,\lambda}$ of the mode frequencies $\omega_\bk, \tilde{\omega}_{\bq,\lambda}$ contribute.
In this limit, and assuming that the occupation factors $n(\omega)$ and damping rates $\eta_\bk,\tilde \eta_{\bq,\lambda}$ are approximately constant for the narrow frequency range mentioned above, we can approximate these rates with the expressions given in Eq.~\eqref{eq:FullGammas}. 

\bibliography{References}

\end{document}